\documentclass{cs20proc}

\editors{S. J. Wolk}
\publisher{Zenodo}
\conference{The 20th Cambridge Workshop on Cool Stars, Stellar Systems, and the Sun}
\conferencedate{2018}

\title{Observations of the solar chromosphere with ALMA and comparison with theoretical models}

\author{R. Braj\v sa,$^{1}$ 
	   D. Sudar,$^{1}$
	I. Skoki\'c,$^{1}$
	A. O. Benz,$^{2,3}$
	M. Kuhar,$^{2,3}$
	A. Kobelski,$^{4}$
    S. Wedemeyer,$^{5}$ 
    S. M. White,$^{6}$
	H.-G. Ludwig,$^{7}$
	M. Temmer,$^{8}$
	S. H. Saar,$^{9}$
	C. L. Selhorst$^{10}$ 
			      }

\affiliation{$^{1}$ Hvar Observatory, Faculty of Geodesy, University of Zagreb, 10000 Zagreb, Croatia\\
		$^{2}$ University of Applied Sciences and Arts Northwestern Switzerland, 5210 Windisch, Switzerland\\ 
		$^{3}$ Institute for Particle Physics and Astrophysics, ETH Zurich, 8093 Zurich, Switzerland\\
		$^{4}$ Department of Physics and Astronomy, West Virginia University, Morgantown, WV 26506-6315, USA\\
		$^{5}$ Institute of Theoretical Astrophysics, University of Oslo, 0315 Oslo, Norway\\		
		$^{6}$ Space Vehicles Directorate, AFRL, Kirtland AFB, NM 87117-5776, USA\\
		$^{7}$ ZAH, Landessternwarte Königstuhl, 69117 Heidelberg, Germany\\
		$^{8}$ Institute of Physics, University of Graz, 8010 Graz, Austria\\		
		$^{9}$ Harvard-Smithsonian Center for Astrophysics, Cambridge, MA 02138, USA\\
		$^{10}$ NAT - Nucleo de Astrofisica Teorica, Universidade Cruzeiro do Sul, Sao Paulo, SP, Brazil
		}

\shorttitle{Solar Observations with ALMA}
\shortauthors{R. Braj\v sa et al.}

\abs{In this work we use solar observations with the ALMA radio telescope at the wavelength of 1.21 mm. 
The aim of the analysis is to improve understanding of the solar chromosphere, a dynamic layer in the solar 
atmosphere between the photosphere and corona.  
The study has an observational and a modeling part. In the observational part full-disc solar images are analyzed. 
Based on a modified FAL atmospheric model, radiation models for various observed solar structures are developed. 
Finally, the observational and modeling results are compared and discussed.  
}

\begin{document}

\maketitle

\section{Introduction}

The Atacama Large Millimetre/submillimetre Array
(ALMA)\footnote{http://www.almaobservatory.org}
is currently the world largest ground-based astronomical facility,
capable of observing almost all types of 
celestial objects including the Sun \citep{Kobelski2016, Barta2017, Bastian2018}.
The main advantage of solar observations with ALMA is mapping of the solar 
chromosphere with an unprecedented spatial, temporal, and spectral resolution
in the  wavelength range between 0.3 mm and 8.6 mm
\citep{Wedemeyer2016, Shimojo2017, White2017}. 
Solar measurements are currently limited to two observing bands centered at 
1.3 mm (239 GHz, Band 6) and 3 mm (100 GHz, Band 3) \citep{Brajsa2018}. 

A valuable and unique property of solar ALMA measurements is 
its capability to be used as an approximately 
linear thermometer of plasma in the solar atmosphere \citep{Wedemeyer2016}. 
So, the measured brightness 
temperature (the intensity of radiation) is directly proportional to the gas temperature of the observed structure or 
layer in the solar atmosphere.
The formation height of the continuum radiation increases with increasing observing wavelength which enables very 
accurate measurements of the solar chromosphere's temperature as a function of height. 
The topic is important for solar physics, but it is important for stellar physics too, since the Sun is representative for the whole class of solar-like 
and other late-type stars \citep{Aschwanden2008, Liseau2016}. 

This study has an observational and a modeling part. 
In the observational part, data reduction is performed on 
Commissioning and Science Verification (CSV) data taken 
during several test campaigns in previous years and made publicly available in 2017. 
Models of various observed solar structures were developed and compared with actual ALMA observations. 
Radiation models are based on modified FAL atmosphere models with thermal bremsstrahlung as the dominant mechanism responsible for the emission at ALMA wavelengths. 
A comparison of observations and models enables precise constraints on plasma properties in the solar atmosphere.

\section{Solar ALMA observations and measurements of the brightness temperature}

In present analysis we use fast-scan single-dish mapping of the Sun. The observing method, calibration, calculation of the 
brightness temperature and producing full-disc solar images are described in detail by \citet{White2017}. 

Many Commissioning and Science Verification (CSV) data of solar observations with ALMA were released to the 
scientific community in 2017\footnote{https://almascience.eso.org/alma-data/science-verification}.
We have used an image of the whole solar disc from December 18th, 2015 taken with a 
12 m single dish total power ALMA antenna at a frequency of 248 GHz corresponding to 
$\lambda = 1.21$ mm in a double circle pattern \citep{Brajsa2018}. 
The measurement frequency/wavelength is in  Band 6.
The beam size amounts to 26 arcsec and the ALMA solar map used here is presented in Fig. 1 in the paper 
by \citet{Brajsa2018} compared with corresponding full-disc solar images in EUV (AIA-SDO data) and H$\alpha$ and 
with the solar magnetogram (HMI-SDO data). 

In the full-disc solar ALMA image at 1.21 mm (Fig. 1a in \citet{Brajsa2018}) several regions of interest were 
identified. They correspond to the quiet Sun areas, active regions, sunspots, filaments (prominences on solar disc), 
magnetic inversion lines and coronal holes. The areas of typical representatives of those solar structures were reduced, 
so that their average brightness temperature could be measured. 

We performed a qualitative and a quantitative analysis of the solar ALMA image. Qualitatively we concluded 
that at 1.21 mm active regions are bright, but a sunspot within the active region appears dark. The inversion lines 
of the large scale magnetic field are dark at 1.21 mm and the ALMA structures outline also the shape of these 
objects. Finally, filaments and coronal holes can be barely visually discerned from the quiet Sun background. 

Quantitatively we found following results: the brightness temperature of the quiet Sun region at the solar disc 
centre was measured to be 6040 K at 1.21 mm. This is fully consistent with the central brightness temperature 
determined by \citet{White2017} for the 2015 data: $T_b = 6040 \pm 250$ K. Further, a limb brightening of quiet 
Sun regions, up to 10\% was measured, in general agreement with some previous results for this wavelength range. 
The measured active region (Table 1 in \citet{Brajsa2018}) had a higher brightness temperature than 
surrounding areas by about 1000 K. In the sunspot 90 K depression relative to the quiet Sun intensity at the 
same radial distance from the solar disc centre was measured. The magnetic inversion lines, filaments and coronal holes all 
had smaller brightness temperatures compared to the quiet Sun levels at their corresponding radial distances from the 
disc centre. Their relative intensities were: $\Delta T_b = - 170$ K (magnetic inversion line), $\Delta T_b = - 110$ K 
(filament), and $\Delta T_b = - 50$ K (coronal hole). 

Finally, we studied the small bright isolated structures in the ALMA 1.21. mm solar image \citep{Brajsa2018}. 
These are the so called 
ALMA bright points and they have a high overlapping rate with coronal bright points (AIA-SDO 19.3 nm), He I 1083 nm 
dark points (NSO-SOLIS), and small-scale magnetic features (SDO-HMI). 
Presently we do not have a quantitative analysis of ALMA bright points and this work is in progress. 

\section{Modeling of the brightness temperature for various solar structures}

Thermal bremsstrahlung is the dominant radiation mechanism assumed in present analysis. 
With decreasing wavelengths the optical depth $\tau = 1$ is reached at 
lower heights in the solar atmosphere with lower temperatures. 
Our investigation is based on the models of \citet{Fontenla1993}, describing average models that agree reasonably
well with radio observations \citep{Bastian1996}. 
The average model is then disturbed to find the necessary deviations for the observed structures yielding the 
physical parameters differing from the characteristics of the quiet solar atmosphere. 

The calculation was performed using a program that computes the brightness temperature for 
a defined wavelength
and stores the increase in brightness temperature per unit height in an array. Then,  the program integrates
these contributions and yields the total brightness temperature for the given 
wavelength. Finally, the procedure is repeated for all wavelengths under consideration. 

Our starting model of the solar chromosphere and corona is the {\em FAL model A} \citep{Fontenla1993},
combined with the Baumbach-Allen coronal model at high altitudes using an electron temperature
$T_{e}=1.2\cdot 10^{6}$ K \citep{Benz1997}. This model roughly describes the structure of a coronal hole
\citep{Brajsa2007}, and we use it in the present analysis as a model of the quiet Sun (Model QS).

\subsection{Active regions}

The higher brightness temperature in active regions is primarily a consequence of enhanced density in the chromosphere and corona.
This shifts the $\tau = 1$ point to higher altitudes where the temperature is higher. 
Above active regions, at the heights from 100 km to 80 000 km, the solar atmosphere has density and
temperature values which are different from those above the quiet Sun regions. 
Three active region models were developed and in all three cases the temperature is higher by factor of 2, 
while the density is larger by the factors of 5, 7, and 10.
All three active region models have significantly higher brightness temperatures than the quiet Sun in the whole 
ALMA wavelength range. Thus, active regions should  appear very bright at 
ALMA wavelengths. 

The preliminary results are presented in Fig. 2b in the paper by  \citet{Skokic2017}. 
We can see that the brightness temperature of all three active region models at the wavelength of 
about 1 mm are higher by about 7000 K than the quiet Sun level.

\subsection{Prominences on solar disc (filaments)}

Taking into account the physical parameters of prominences \citep{Engvold1990, 
Tandberg1995} we develop six prominence models. 
Prominences are denser and cooler structures in the solar atmosphere. 
The prominence models assume hydrostatic equilibrium and thus pressure is conserved at a given altitude. 
So, the values of the density, $n$, are multiplied by a factor $f$ and 
the values of the temperature, $T$, are divided by the same factor $f$ \citep{Parenti2014}. 
This factor $f$ amounts to $f$=80, $f$=120, $f$=160, $f$=200, $f$=240, $f$=280,
for the six  prominence models, at the prominence heights from 40 000 km to 50 000 km, 
which are typical prominence heights, see, e.g., \citet{Bastian1993} and \citet{Brajsa2009}. 

The calculated brightness temperatures for the six prominence models 
are presented as a function of wavelength in Fig. 3a in the paper by  \citet{Skokic2017}. 
We see that for some models there are two radiation regimes: absorption and emission, dependent on the wavelength. 
At some specific wavelength, depending on the model, filaments become invisible against the 
background radiation of the quiet Sun and a transition from absorption to emission takes place. 
At the wavelength of about 1 mm the curves (the brightness temperature vs. wavelength) converge 
and some models predict a small excess in intensity, while one model remains slightly below the quiet Sun 
level. 

\subsection{Coronal holes}

Our starting model of the quiet Sun describes the conditions similar to the 
coronal hole atmosphere.  
We now construct deviations from the coronal hole model  by 
changing the values of the density and temperature towards the structure of the quiet non-hole chromosphere and 
corona. These models which simulate various non-hole structures will be referred to 
as non-hole models.  
We develop four quiet non-hole solar atmosphere models taking into account 
that coronal holes are regions of lower temperature and density in the solar 
corona. 
For the non-hole atmosphere, 
the hybrid network model of \citet{Gabriel1992}  is used. 
Temperature and density parameters used for constructing these models are 
based on various studies from the literature. 

The resulting brightness temperatures for all models (coronal hole and 4 non-hole models) are shown 
in Fig. 3b in the paper by  \citet{Skokic2017}. 
It can easily be seen that there is no significant difference in predicted intensity between the quiet Sun and 
coronal holes for the main ALMA wavelength range (wavelengths from 3 mm to 
0.3 mm, corresponding to bands 3 to 10). The difference becomes smaller with decreasing wavelength and 
the curves converge in the mm wavelength range.

Finally, we note that the modeling part of this work is performed within the SSALMON\footnote{https://www.ssalmon.uio.no} 
international scientific network \citep{Wedemeyer2015a, Wedemeyer2015b}.
Modeling efforts so far, relevant for the present analysis, are described in a preliminary form 
by \citet{Skokic2017} and further work is currently underway.

\section{A comparison of observational and theoretical results, discussion and concluding remarks}

In the present analysis we calculated also the brightness temperature of the quiet Sun for the ALMA wavelength 
range, but we note that  the determination of the absolute quiet Sun level is not an easy task, both theoretically and 
experimentally \citep{White2017}. So, we will make a comparison between the observed and calculated results only 
relatively, as differences from the quiet Sun level. We now summarize these results for the three analyzed structures 
in the solar atmosphere: active regions, prominences on the disk and coronal holes, for the wavelength of 1.21 mm. 

The observed active region had a brightness temperature higher than the quiet Sun level by about 1000 K. However, 
the calculated value is much higher, about 7000 K. This indicates a qualitatively correct result, but of a much larger 
calculated value than the measured one. The model could be improved by taking smaller values of temperature and 
density increase and by checking how the range of integration (which corresponds to the height and position of 
the active region in the solar atmosphere) influences the calculated brightness temperature.  

For prominences on the disc the measured value indicates a slight absorption (a negative difference to the quiet Sun 
level of about 100 K) which is consistent with only one (out of six) prominence models with the highest factor for 
increasing the density and decreasing the temperature. One possible interpretation is that the observed prominence 
has the parameters which put into the model reproduce the measurements in the best way. The possibility that 
other prominences would behave differently can not be excluded and further observations are needed. 

Coronal holes have a very small negative difference between the measured values inside the hole and the 
surrounding quiet Sun areas. This is fully consistent with theory as we have seen that all models (coronal hole 
and non-hole) converge at mm wavelengths. It should also be noted that it is not always trivial to determine the borders 
of coronal holes from observations, in spite of huge improvement of modern detection techniques. Special care 
should be taken to avoid possible misidentifications. Moreover, coronal holes are rarely fully homogeneous structures. 
Small localized brightennings often appear within them \citep{Brajsa2007, Selhorst2017} which should be taken into account in ALMA image 
analysis and comparison with other data, as well as in modeling efforts. 

In present work we have described observational and modeling efforts to reconstruct and interpret full-disc solar images 
at the wavelength of 1.21 mm recorded with ALMA. The work will be continued by including other observing bands, adding 
the interferometric analysis and refining theoretical models.

\section*{Acknowledgments}
{R.B. acknowledges partial support from the scientific grant of the University of Zagreb. 
This work has been supported in part by Croatian Science Foundation under the project no. 7549 
"Millimeter and submillimeter observations of the solar chromosphere with ALMA" (MSOC). Also, 
we acknowledge the Austrian-Croatian bilateral scientific project, funded by the Ministries of Science of 
Austria and Croatia in 2018-2019. 
S.W. is supported by the SolarALMA project, which has received funding from the European Research Council (ERC) under the European 
Union's Horizon 2020 research and innovation programme (grant agreement No. 682462), and by the Research Council of Norway through its Centres of Excellence scheme, project number 262622.
This paper makes use of the following ALMA data: ADS/JAO.ALMA\#2011.0.00020.SV.
ALMA is a partnership of ESO (representing its member states), NSF (USA) and
NINS (Japan), together with NRC (Canada) and NSC and ASIAA (Taiwan), and KASI
(Republic of Korea), in cooperation with the Republic of Chile. The Joint ALMA
Observatory is operated by ESO, AUI/NRAO and NAOJ.
We are grateful to the ALMA project for making solar observing with ALMA possible.
SDO is the first mission launched for NASA's Living With a Star (LWS) Program.
This work utilizes GONG data obtained by the NSO Integrated Synoptic Program (NISP), 
managed by the National Solar Observatory, the Association of Universities for Research in 
Astronomy (AURA), Inc. under a cooperative agreement with the National Science Foundation. 
The data were acquired by instrument operated by the Cerro Tololo Interamerican Observatory.
The National Radio Astronomy Observatory is a facility of the National 
Science Foundation operated under cooperative agreement by Associated 
Universities, Inc.
}

\bibliographystyle{cs20proc}
\bibliography{Brajsa_1.bib}

\end{document}